\documentclass[prd,aps,showpacs,nofootinbib,nobibnotes,superscriptaddress]{revtex4}

\usepackage{epsfig}
\usepackage{amssymb}
\usepackage{graphics}
\usepackage{bm}

\begin{document}

\draft

\title{Gravitational Radiation from Primordial Helical  Inverse Cascade MHD Turbulence}

\date{\today}

%\date{\today}

\author{Tina Kahniashvili}
%\email{tinatin@phys.ksu.edu}
\affiliation{Department of Physics,
Kansas State University, 116 Cardwell Hall, Manhattan, KS 66506,
USA} \affiliation{Department of Physics, Laurentian University,
Ramsey Lake Road, Sudbury, ON P3E 2C6, Canada}
\affiliation{National Astrophysical Observatory, Ilia
Chavchavadze State University, 2A Kazbegi Ave, Tbilisi, GE-0160,
Georgia}

\author{Leonardo Campanelli}
%\email{leonardo.campanelli@ba.infn.it}
\affiliation{Dipartimento
di Fisica, Universit\`{a} di Bari, I-70126 Bari, Italy}
\affiliation{INFN - Sezione di Bari, I-70126 Bari, Italy}

\author{Grigol Gogoberidze}
%\email{gogober@geo.net.ge}
\affiliation{Centre for Plasma Astrophysics, K.U.~Leuven,
Celestijnenlaan 200B, 3001 Leuven, Belgium}
\affiliation{Department of Physics, Kansas State University, 116
Cardwell Hall, Manhattan, KS 66506, USA} \affiliation{National
 Astrophysical Observatory, Ilia Chavchavadze State University, 2A
Kazbegi Ave, Tbilisi, GE-0160, Georgia}

\author{Yurii Maravin}
%\email{maravin@phys.ksu.edu}
\affiliation{Department of Physics,
Kansas State University, 116 Cardwell Hall, Manhattan, KS 66506,
USA}

\author{Bharat Ratra}
%\email{ratra@phys.ksu.edu}
\affiliation{Department of Physics, Kansas State University, 116 Cardwell Hall, Manhattan, KS 66506, USA}

\begin{abstract}
We consider the generation of gravitational waves by primordial
 helical inverse cascade magnetohydrodynamic (MHD) turbulence
produced by bubble collisions at the electroweak phase
transition. We extend the previous study~\cite{kgr08} by
considering both currently discussed  models of MHD turbulence.
For popular electroweak phase transition parameter values, the
generated gravitational wave spectrum is only weakly dependent on
the MHD turbulence model. Compared to the unmagnetized
electroweak phase transition case, the spectrum of
MHD-turbulence-generated gravitational waves peaks at lower
frequency with larger amplitude and can be detected by the
proposed Laser Interferometer Space Antenna.
\end{abstract}

\pacs{98.70.Vc, 98.80.-k}

\maketitle

\section{introduction}

The direct detection of a gravitational wave background will
provide us with a probe of physical conditions in the early
Universe at the epoch when the gravitational radiation was
generated. This is because gravitational waves are massless and
after generation weakly-coupled gravitational radiation
propagates freely. Hence, once generated, any gravitational wave
spectrum retains its shape, with all wavelengths simply scaling
with the expansion of the Universe. Gravitational wave astronomy
holds the potential for helping construct a picture of the
Universe at energy scales even higher than those the Large Hadron
Collider will reach, to probe in detail physics at the
electroweak energy scale.

There are various mechanisms that might generate gravitational
waves in the early Universe, for reviews see~\cite{source,m00}.
A well-known one is parametric amplification of quantum
fluctuations~\cite{inflation}, which can also take place shortly after
inflation~\cite{shortafter}.  Other mechanisms include bubble wall motion
and collisions during phase transitions; gravitational wave generation
during the
electroweak and QCD phase transitions are discussed in
Refs.~\cite{bubble,kos1,huber,nicolis,cds06}. Cosmic strings and
other defects can produce relic gravitational waves~\cite{strings}.
Cosmological magnetic fields~\cite{magnet,cdk04,cd06} and
hydrodynamical or magnetohydrodynamical turbulence can also induce
primordial gravitational waves~\cite{kmk02,dolgov,kgr05,gkk07,kgr08}.

Gravitational waves will also provide information on symmetries
present in the Universe. In particular, the relic gravitational
wave background carries information about parity asymmetry, that
might be generated at the electroweak phase transition, or
earlier. Parity symmetry testing based on the direct detection of
gravitational waves in the near future (with currently planned
missions) seems to be promising for gravitational waves with
frequencies around the Laser Interferometer Space Antenna (LISA)
sensitivity band at $0.1 - 100$ mHz~\cite{lisa}. The Standard
Model electroweak phase transition, when parity symmetry breaking
and magnetic helicity production might be
expected~\cite{helicity}, produces gravitational waves with a
significant amplitude in this frequency range. Recently, some of
us have investigated the generation of gravitational waves
through inverse cascade MHD turbulence~\cite{kgr08} and found
that even a small amount of initial helicity enlarges the range
of phase transition parameters for which gravitational waves are
potentially detectable by LISA. This gravitational wave signal
has an important feature: it is circularly polarized. For direct
cascade hydrodynamical turbulence some of us have previously
estimated the polarization degree~\cite{kgr05}, which is obviously
turbulence model dependent. Polarized gravitational waves are
present in other models~\cite{gw-pol}, and the polarization of
the gravitational wave background is in principle observable,
either directly~\cite{seto} or through the cosmic microwave
background~\cite{cdk04,a}.

The detection prospects of gravitational radiation from the early
Universe depend on the energy scale when gravitational waves were
generated and even more  so on the source duration time and
efficiency of the generation process itself. In particular, to
produce a detectable signal the electroweak phase transition must
be strong enough~\cite{detection,gs06,grojean}. The currently
popular standard model of particle physics does not have a strong
enough electroweak phase transition. Also, all currently discussed
electroweak phase transition models do not predict an observable
gravitational radiation signal if one assumes only bubble
collisions~\cite{grojean}. It is of particular interest to study
gravitational wave production during phase transitions in the minimal
and next to minimal supersymmetric standard models (MSSM and nMSSM
\cite{models}), \cite{huber}. The gravitational wave signal produced
in the MSSM model is below the LISA sensitivity, while the nMSSM
results in a stronger electroweak phase transition \cite{models}
and as a consequence a detectable  gravitational wave signal
\cite{huber}.

On the other hand, sources associated with stochastic vector fields
in plasma, i.e.\ kinetic velocity or magnetic fields, significantly
change gravitational wave detection prospects~\cite{kkgm08}.  In
the case of unmagnetized hydrodynamical turbulence the peak frequency
of the gravitational wave power spectrum is determined by the
characteristic time of turbulence, i.e. the inverse turn-over time
of the largest eddy. Another important characteristic is the
energy scale when gravitational radiation is generated. In recent
modifications of the standard electroweak
phase transition model where the transition is moved to a higher
energy scale~\cite{gs06,rs06}, the gravitational wave power
spectrum peak frequency is shifted to a higher frequency,
which, since the gravitational wave spectrum is sharply peaked,
reduces the possibility of detection by LISA. The presence of a
cosmological magnetic field affects the dynamics of turbulence.
In the case of MHD turbulence the presence of an energy
inverse-cascade leads to an increase in the effective size of the
largest eddy (now associated with a helical magnetic field), and
can result in the gravitational wave power spectrum peaking in
the LISA band, with amplitude large enough to be detected by
LISA~\cite{kgr08}.

In this paper we extend the study of Ref.~\cite{kgr08} by also
considering a different model of MHD turbulence. In particular,
Ref.~\cite{kgr08} adopts the inverse cascade model of
Refs.~\cite{BM99,CHB05}, and in this paper we also include in the
consideration the model of Refs.~\cite{son,jedamzik,campanelli}.
In both cases,  gravitational wave generation is considered
assuming non-zero magnetic helicity during the phase transition.

For our analysis in this paper we adapt the technique developed
in Ref.~\cite{gkk07}. We model MHD turbulence and obtain the
gravitational wave spectrum by using an analogy with the theory
of sound wave production by hydrodynamical
turbulence~\cite{L52,P52,G,my75}. We perform the computation of
the gravitational wave spectrum in real space, instead of using
conventional Fourier space techniques as in
Refs.~\cite{kmk02,cds06}. This makes the physical interpretation
of all quantities straightforward. We use natural units with
$\hbar = c = k_B = 1$ throughout.

The paper is organized as follows. In Sec.~II we summarize the
general formalism of gravitational wave generation. In Sec.~III we
describe the turbulence models we consider. In Sec.\ IV we derive
the gravitational wave spectra, and we conclude in Sec.\ V.

\section{Gravitational Wave Generation Formalism}

We assume that primordial turbulence is generated at  time
$t_\star$ --- corresponding to temperature $T_\star$ at  the phase
transition --- at characteristic proper length-scale $l_0 =
2\pi/k_0$ (where $k_0$ is the corresponding proper wavenumber) and
with characteristic velocity perturbation $v_0$. We assume that
the kinetic and magnetic Reynolds numbers are much greater than
unity on scales of order  $l_0$, otherwise there is no
turbulence. We also assume that the duration of the turbulence,
$\tau_T$, is no larger than the Hubble time at the phase
transition $H^{-1}_\star$, where $H_\star$ is the value of the
Hubble parameter at the phase transition, so the expansion of the
Universe may be neglected during the generation of gravitational
radiation~\cite{kmk02,dolgov}. This adiabatic assumption will be
valid for any turbulence produced in a realistic cosmological
phase transition~\cite{tww91}. Therefore, the gravitational
radiation equation in real space can be written as~\cite{W}
\begin{equation}
\nabla^2 h_{ij}({\mathbf x}, t) - \frac{\partial^2}{\partial t^2}
h_{ij}({\mathbf x}, t) = -16\pi G S_{ij}({\mathbf x}, t),
\label{eq:01}
\end{equation}
where $h_{ij}({\mathbf x}, t)$ is the tensor metric perturbation,
$t$ is physical time, $i$ and $j$ are spatial indices (repeated
indices are summed), and $G$ is the gravitational constant.
$S_{ij}$ is the traceless part of the stress-energy tensor $T_{ij}
({\mathbf x}, t)$, (constructed from either kinetic or magnetic
turbulence vector fields), given by~\cite{W}
\begin{equation}
S_{ij}({\mathbf x}, t) = T_{ij}({\mathbf x}, t) - \frac{1}{3} \,
\delta_{ij} T^k_k({\mathbf x}, t). \label{eq:02}
\end{equation}

Since the turbulent fluctuations are stochastic, so are the
generated gravitational waves. Our goal is to derive the energy
density spectrum of these gravitational wave perturbations at the
end of the turbulent phase; after that the amplitude and
wavelength of the gravitational radiation scales simply with the
expansion of the Universe. The energy density of gravitational
waves is~\cite{m00}
\begin{equation}
\rho_{GW}({\bf x},t)= \frac{1}{32\pi G} \langle
\partial_t h_{ij}({\mathbf x},t) \partial_t h_{ij} ({\mathbf x},t)\rangle
= \frac{G}{2\pi } \int {\rm d}^3 {\bf x}^\prime {\rm d}^3 {\bf
x}^{\prime \prime} \frac{\langle \partial_t S_{ij}({\mathbf
x}^\prime,t^\prime) \partial_t S_{ij}({\mathbf x}^{\prime
\prime},t^{\prime \prime}) \rangle } {|{\bf x}- {\bf x}^\prime|
|{\bf x}- {\bf x}^{\prime \prime}|} \, , \label{eq:04}
\end{equation}
where the brackets denote an ensemble average over realization of
the stochastic source, and the times $t^\prime = t-|{\bf x} - {\bf
x}^\prime|$, $t^{\prime \prime} = t - |{\bf x} - {\bf
x}^{\prime\prime}|$.

We consider metric perturbations in the far-field limit (i.e.\ for
$x \gg d$, where $d$ is a characteristic length scale of the
source region), where gravitational waves are the only metric
perturbations~\cite{W}, and  replace $|{\bf x} - {\bf x}^\prime|$
by $|\bf x|$ in Eq.~(\ref{eq:04}). For the stationary turbulence
we consider here,  we define the gravitational wave spectral
energy density, $I({\bf x},\omega)$, as the one-dimensional
temporal Fourier transform of the autocorrelation function of the
temporal derivative of the  tensor metric perturbations
$L(\bf{x},\tau)$,
\begin{equation}
I({\bf x},\omega) = \frac{1}{2\pi} \int {\rm d} \tau e^{i\omega
\tau} L({\bf x},\tau), \label{eq:08}
\end{equation}
where $\omega$ is the angular frequency and
\begin{equation}
L({\bf x},\tau) = \frac{1}{32\pi G} \, \langle
\partial_t h_{ij}({\mathbf x},t) \partial_t h_{ij} ({\mathbf x},t+\tau)
\rangle.
\label{Ldef}
\end{equation}
Accounting for Eq.~(\ref{eq:04}), the gravitational wave energy
density is
\begin{equation}
\rho_{\rm GW}({\bf x}) = \int {\rm d} \omega~I({\bf x},\omega).
\label{spectral} \end{equation}
Following Ref.~\cite{gkk07} it can be shown that
\begin{equation}
I({\bf x},\omega) = \frac{4\pi^2\omega^2 G {\rm w}^2}{|{\bf x}|^2}
\int {\rm d}^3 {\bf x}^\prime H_{ijij} \left( {\bf x}^\prime,
\frac{{\bf x}}{|{\bf x}|} \omega, \omega\right), \label{eq:20}
\end{equation}
where ${\bf k}$ is a proper wavevector and ${\rm w} = \rho + p$,
with ${\rm w}$, $\rho$, and $p$ being respectively the enthalpy,
pressure, and energy density of the plasma. The quantity
$H_{ijij}({\bf x}^\prime, {\bf k}, \omega)$ is the double trace
of the  four-dimensional temporal and spatial Fourier
transform\footnote{Consistent with current observational
indications \cite{rv08}, we assume  flat spatial hypersurfaces and
 use spatial Fourier transform relations $F_j({\mathbf k}) =
\int d^3\!x \,
   e^{i{\mathbf k}\cdot {\mathbf x}} F_j({\mathbf x})$ and
   $F_j({\mathbf x}) = \int d^3 k~
   e^{-i{\mathbf k}\cdot {\mathbf x}} F_j({\mathbf k})/(2\pi)^3$.}, taken
with respect to ${\bf x}^{\prime \prime} - {\bf x}^{\prime}$ and
$\tau$, of the two-point time-delayed fourth-order correlation
tensor
\begin{equation}
R_{ijkl}({\bf x}^\prime, {\bm \xi}, \tau) = \frac{1}{{\rm w}^2} \,
\langle S_{ij}({\bf x}^\prime ,t) S_{kl}({\bf x}^{\prime \prime},
t+\tau) \rangle.
\end{equation}
That is,
\begin{equation}
H_{ijkl}({\bf x}^\prime,{\bf k},\omega) = \frac{1}{(2\pi)^4} \int
{\rm d}^3 {\bm \xi} ~{\rm d} \tau~ e^{i(\omega \tau - {\bf k}\cdot
{\bm \xi})} R_{ijkl}({\bf x}^\prime, {\bm\xi}, \tau),
\label{hijkldef}
\end{equation}
where ${\bm \xi} = {\bf x}^{\prime \prime}-{\bf x}^\prime$.

Given the function $H_{ijkl}({\bf x}^\prime,{\bf k},\omega)$, we
may use Eqs.\ (\ref{spectral}) and (\ref{eq:20})
 to compute the gravitational wave
energy density, assuming that the source is statistically
homogeneous, so that the averaged correlators of the stress-energy
tensor have no spatial dependence, and isotropic, so that the
correlation between two spatial points depends only on the
distance between the points and not on the direction. With these
assumptions, Eq.~(\ref{hijkldef}) reduces to
\begin{equation}
H_{ijkl}({\bf x}^\prime,{\bf k},\omega) = H_{ijkl}({\bf k},\omega)
= \frac{1}{(2\pi)^4} \int {\rm d}^3 {\bm \xi}~ {\rm d} \tau~
e^{i(\omega \tau - {\bf k}\cdot {\bm \xi})} R_{ijkl}({\bm\xi},
\tau),
\end{equation}
so $H_{ijkl}(\hat{{\bf x}} \omega,\omega) =
H_{ijkl}(\omega,\omega)$ is independent of the observation
direction $\hat{{\bf x}}$, as expected on physical grounds.

Finally, since the gravitational waves propagate freely after been
generated, the expansion of the Universe is accounted for by a
simple re-scaling of the frequency and the amplitude by a factor
equal to
\begin{equation}
\frac{a_\star}{a_0} \simeq 8 \times 10^{-16} \left(\frac{100\,{\rm
GeV}}{T_\star}\right) \left(\frac{100}{g_\star}\right)^{{1}/{3}},
\label{a-ratio}
\end{equation}
where $g_\star$ is the number of relativistic degrees of freedom
at the temperature $T_\star$. For the standard model degrees of
freedom we have $g_*= 106.75$ as $T \rightarrow \infty$. With
these assumptions and considering a stochastic turbulence source
lasting for a finite duration $\tau_T$, the total gravitational
radiation energy density spectrum at a spatial point and a time
can be obtained by integrating over all sources within a spherical
shell centered at that observer, with a shell thickness
corresponding to the duration of the phase transition, and a
radius equal to the proper distance along any light-like path
from the observer to the source. Then we get~\cite{gkk07,kgr08}
\begin{equation}
\rho_{GW}(\omega) = \frac{d\rho_{GW}}{d\ln\omega}= 16\pi^3\omega^3
G \, {\rm w}^2 \tau_T H_{ijij}(\omega,\omega), \label{spectrum}
\end{equation}
where $\omega$ is the angular frequency measured at the moment of
generation of the gravitational radiation. As shown in
Ref.~\cite{gkk07}, while $H_{ijij}(\omega,\omega)$ is a
complicated function of $\omega$, the resulting gravitational wave
signal strength can be simply estimated with $\pm 25\%$ accuracy
by working in the  ${\bf k} \rightarrow 0$ limit and  replacing
$H_{ijij}({\bf k},\omega)$ by the corresponding
$H_{ijij}(0,\omega)$  --- this is the main point  of the
aero-acoustic approximation initially used for sound wave
generation by turbulence~\cite{G}. It can be shown that this
approximation works rather well for low velocity turbulence, with
Mach number $M$ less than unity, i.e.\ for $M = v_0 < 1$, where
$v_0$ corresponds to the turbulence velocity of the largest eddy.

We assume that turbulence is caused by bubble collisions during
the phase transition. In general, phase transitions can be
described by two key parameters: $\alpha = \rho_{\rm vac} /
\rho_{\rm thermal}$, the ratio of the vacuum energy density
associated with the phase transition to the thermal energy
density of the Universe at the time (this characterizes the
strength of the phase transition); and $\beta^{-1}$, which sets
the characteristic time scale for the phase transition, as well
as the temperature  $T_\star$ when the phase transition occurs.
All other parameters, such as the efficiency factor $\kappa$ which
gives the fraction of the available vacuum energy which goes into
the kinetic energy of the expanding bubble walls (as opposed to
thermal energy), the bubble wall velocity $v_b$, and the turbulent
eddy velocity $v_0$, can be expressed through $T_\star$,
$\alpha$, and $\beta^{-1}$. In
particular~\cite{kos1,ste82,nicolis}
\begin{eqnarray}
\kappa(\alpha) &=& {1\over 1+0.715 \alpha} \left[0.715 \alpha +
{4\over
27}\left(3\alpha\over 2\right)^{1/2}\right] \! , \label{kappa} \\
v_b(\alpha) &=& {1/\sqrt{3} + (\alpha^2 + 2\alpha/3)^{1/2} \over
1 + \alpha} \, , \label{vb} \\
v_0 (\alpha) &=& \left(\frac{3\kappa\alpha}{4 + 3\kappa \alpha}
\right)^{\!1/2}. \label{v0}
\end{eqnarray}
 The parameter $\beta^{-1}$ is determined from  the bubble nucleation rate per unit volume $ \Gamma = \Gamma_0 e^{\beta t}$ \cite{kos1} where $\Gamma_0$
 is determined by initial conditions. For the standard model
$\beta^{-1}=100H_\star$. The temperature $T_\star$ and the Hubble
parameter at the phase transition $H_\star$ are related through
\begin{equation}
H^2_* = \frac{8\pi^3 G T_*^4}{90}. \label{H-star}
\end{equation}

In the next section we discuss two specific models for the
turbulent motions.

\section{Turbulence Models}

In this paper we are interested in the generation of
gravitational waves by helical MHD turbulence, that is turbulence
in  a magnetized plasma with non-vanishing magnetic helicity,
${\mathcal H}_M$, defined by
\begin{equation}
{\mathcal H}_M(t) = \int d^3 x~{\bf A} \cdot {\bf B},
\end{equation}
where ${\bf A}$ is the vector potential of the magnetic field
${\bf B}$. Magnetic helicity is  odd  under discrete $P$ and $CP$
transformations, so  the presence of magnetic helicity in our
Universe would be an indication  of  macroscopic $P$ and $CP$
violation.

While MHD turbulence is isotropic on large scales, it is locally
anisotropic on small scales~\cite{s83}, resulting in small-scale
anisotropy in the generated GW background. However, GWs are
generated mainly by the largest eddies~\cite{gkk07} so we adopt an
isotropic turbulence model with velocity field and magnetic field
two-point correlation functions~\cite{my75}
\begin{equation}
\langle v_i^* ({\bf k}, t) v_j({\bf k'}, t+\tau) \rangle =
\delta({\bf k} -{\bf k'}) \, F_{ij}^v({\bf k},t) \,
f[\eta(k),\tau],
\end{equation}
\begin{equation}
\langle b_i^* ({\bf k}, t) b_j({\bf k'}, t+\tau) \rangle =
\delta({\bf k} -{\bf k'}) \, F_{ij}^M\!({\bf k},t) \,
f[\eta(k),\tau]. \label{2-point}
\end{equation}
Here ${\bf v}$ is the turbulent-motion velocity field, ${\bf b} =
{\bf B}/\sqrt{4\pi {\rm w}}$ is the characteristic magnetic field
perturbation ${\bf B}$ expressed in velocity units, and
\begin{equation}
F_{ij}^v({\bf k},\tau) =  P_{ij}({\bf k}) \frac{E_v(k,t)}{4\pi
k^2},\label{20}
\end{equation}
\begin{equation}
F_{ij}^M\!({\bf k},\tau) =  P_{ij}({\bf k}) \frac{E_M(k,t)}{4\pi
k^2} + i \varepsilon_{ijl} {k_l} \frac{H_M(k,t)}{8\pi k^2} \, .
\label{eq:4.1}
\end{equation}
In the above equations the projection operator   $P_{ij}({\bf k})
= \delta_{ij} - {k_i k_j}/{k^2}$, $\delta_{ij}$ is the Kronecker
delta, $k = |{\bf k}|$, $\varepsilon_{ijl}$ is the totally
antisymmetric tensor, and $\eta(k)$ is an autocorrelation
function that determines the characteristic function
$f[\eta(k),\tau]$ describing the temporal decorrelation of
turbulent fluctuations. In the following we use $f[\eta(k),\tau]
= \exp \! \left[-\pi \eta^2(k) \tau^2/4 \right]$~\cite{K64}. The
function $E_v(k,t)$ in Eq.\ (\ref{20}) is the so-called kinetic
power spectrum and is related to the kinetic energy of turbulence
through
\begin{equation}
{\mathcal E}_v(t) = \int \! dk~E_v(k,t),\label{21}
\end{equation}
while in Eq.\ (\ref{eq:4.1}) $E_M(k,t)$ and $H_M(k,t)$, the
magnetic field energy and magnetic helicity power spectra, denote
the symmetric and antisymmetric parts of the two-point magnetic
field correlator. They are related to the magnetic energy and
magnetic helicity densities through
\begin{equation}
{\mathcal E}_M(t) = \int \! dk~E_M(k,t)
\end{equation}
and
\begin{equation}
{\mathcal H}_M(t) = \int \! dk~ H_M(k,t),
\end{equation}
respectively.

It is worth noting that for all magnetic field configurations,
the magnetic helicity spectrum must satisfy the ``realizability
condition''~\cite{B03}:
$|H_M(k,t)| \leq 2 E_M(k,t)/k$.
Introducing the magnetic correlation length (i.e., the
characteristic proper length associated with the large magnetic
energy eddies of turbulence),
\begin{equation}
\xi_M(t) = \frac{\int \! dk k^{-1} E_M(k,t)}{{\mathcal E}_M(t)} \,
,
\end{equation}
the integral form of the realizability condition is %
$|{\mathcal H}_M(t)| \leq 2 \xi_M(t) {\mathcal E}_M(t)$.
In the following, we assume that a magnetic field with (positive)
magnetic helicity has been generated during a phase transition
through some mechanism~\cite{helicity}. We also assume  ``small''
initial magnetic helicity, in the sense that the parameter
\begin{equation}
\zeta_\star = \frac{{\mathcal H}_M(t_\star)}{2 \, \xi_M(t_\star)
{\mathcal E}_M(t_\star)}
\end{equation}
is much smaller then unity.

After generation, primordial helical turbulence  decays freely.
While it is widely accepted in the literature that the free decay
of helical MHD turbulence is a two-stage process, the details of
the two stages are still under discussion. Roughly speaking, in
the first stage, the system proceeds through selective decay of
magnetic modes: magnetic power on small scales is washed out by
turbulence effects more effectively than on large scales. During
this process the magnetic correlation length grows while the
magnetic energy decays in time. The first stage ends when
quasi-conservation of magnetic helicity starts to trigger an
inverse cascade of the magnetic field: small-scale modes are no
longer completely dissipated by turbulence, rather part of their
energy is now transferred to larger scales. This causes a faster
growth of the correlation length and a slower dissipation of the
magnetic energy in comparison to the non-helical case.

Different analyses  in the literature seem to agree on the
details of the first stage of direct cascade. The details of the
inverse-cascade regime, however, are  still being debated  and
there are two main models, that  of Refs.~\cite{BM99,CHB05}, here
referred to as Model A, and that of
Refs.~\cite{jedamzik,campanelli}, which we will call Model B. In
the following, we consider both models.

\subsection{First stage: direct cascade}

In both models the dynamics of the  first decay stage is governed
by a direct cascade of magnetic energy density lasting for a time
$\tau_{s0} = s_0 \tau_0$, a few times ($s_0 \sim 3-5$) longer than
the characteristic largest-eddy turn-over time $\tau_0=l_0/ v_0
=2\pi/k_0 v_0$.~\footnote{To be more precise, from the results of
Refs.~\cite{jedamzik,campanelli}, the time when the system enters
the inverse-cascade regime (in  Model B) could depend on the
fraction of the maximal initial magnetic helicity $\zeta_\star$.
However, those results were obtained assuming an initial magnetic
spectrum of the form $E_M(k,0) \propto k^p$ with $p$ a positive
real number, while in our case the initial spectrum is a red
spectrum of the Kolmogorov type. In the absence of firm results
on this matter, and in order to avoid inessential complications,
here we  assume that in both models of turbulence $\tau_{s0} =
s_0 \tau_0$ with $s_0 \sim $ few.}
At the end of this stage equipartition between kinetic and
magnetic energies is reached~\cite{B03}.

During the first stage magnetic energy density flows from large
to small scales and finally dissipates on the scale $l_d =
2\pi/k_d$ ($k_d \gg k_0$) where one of the Reynolds numbers
becomes of order unity. Due to the selective decay
effect~\cite{B03} magnetic helicity is nearly conserved during
this stage~\cite{CHB05,jedamzik}. To compute the gravitational
waves generated by decaying MHD turbulence, we assume that
decaying turbulence lasting for time $\tau_{s0}$ is equivalent to
stationary turbulence lasting for time $\tau_{s0}/2$. This can be
justified using the Proudman argument for (unmagnetized)
hydrodynamical turbulence~\cite{P52,my75}. Consequently, when
computing the induced gravitational waves we ignore the time
dependence of $E_M(k,t)$ and $H_M(k,t)$. Also, for $E_M(k,t)$ and
$\eta(k)$ we use the Kolmogorov model \cite{kol41} with
\begin{equation}
E_M(k,t) =  C_K \varepsilon^{2/3} k^{-5/3}, \label{eq:4.2}
\end{equation}
and
\begin{equation}
\eta(k) = \frac{\varepsilon^{1/3}}{\sqrt{2\pi}} \: k^{2/3},
\end{equation}
both functions being defined over the range of wavenumbers $k_0 <
k < k_d$. Here $C_K$ is a constant of order unity and
$\varepsilon \simeq k_0 v_0^3$ is the energy dissipation rate per
unit enthalpy. Taking into account that equipartition between
kinetic and magnetic energy densities is established and
maintained during Kolmogorov turbulence,  the total (kinetic plus
magnetic) energy density of turbulence is simply double the
expression in Eq.~(\ref{eq:4.2}).

We note that while turbulence in the early Universe has to be
relativistic, we have assumed a Kolmogorov spectrum for both ${\bf
b}$ and ${\bf v}$, which is valid for  non-relativistic
turbulence. Based on arguments of Ref.~\cite{kmk02}, however, we
expect that our estimate gives the correct qualitative features
of the resulting gravitational radiation spectrum.

\subsection{Second stage: inverse cascade}

\subsubsection{Model A}

Model A has been discussed in Ref.~\cite{kgr08}. Here, we
summarize those results. At the end of the first stage
turbulence relaxes to a maximally helical
state~\cite{CHB05,jedamzik}. Accounting for conservation of
magnetic helicity, the characteristic velocity and magnetic field
perturbations at this stage are $v_1 \simeq \zeta_\star^{1/2}v_0$
and $b_1 \simeq \zeta_\star^{1/2}b_0$. The dynamics of the second
stage is governed by a magnetic helicity inverse cascade. If both
Reynolds numbers are large at the end of the first stage,
magnetic helicity is conserved during the second stage. The
magnetic eddy correlation length evolves as~\cite{BM99,CHB05}
\begin{equation}
\label{xiBM} \xi_M(t) \simeq l_0 \! \left(1 + \frac{t}{\tau_1}
\right)^{\!\!1/2} \! ,
\end{equation}
where $\tau_1 \simeq l_0/v_1 = \tau_0/\zeta_\star^{1/2}$ is the
characteristic eddy turn-over time at the beginning of the second
stage. The magnetic and kinetic energy densities evolve
as~\cite{BM99,CHB05}
\begin{eqnarray}
{\mathcal E}_M(t) & \simeq & {\rm w} b_1^2 \! \left(1 +
\frac{t}{\tau_1} \right)^{\!\!-1/2} \! , \nonumber
\\
{\mathcal E}_v(t) & \simeq & {\rm w} v_1^2 \! \left(1 +
\frac{t}{\tau_1} \right)^{\!\!-1} \! . \label{eq:4.06}
\end{eqnarray}
These imply that the characteristic turn-over ($\tau_{\rm to}$)
and cascade ($\tau_{\rm cas}$) timescales evolve as
\begin{eqnarray}
\tau_{\rm to} \simeq \tau_{\rm cas} \simeq \tau_1 \! \left(1 +
\frac{t}{\tau_1} \right) \! . \label{eq:4.7}
\end{eqnarray}

To compute the gravitational waves emitted during the second stage
we use the stationary turbulence model that has the same
gravitational wave output. Introducing the characteristic proper
wavenumber $k_\xi(t) = 2\pi/\xi_M(t)$ and using
Eq.~(\ref{eq:4.06}) we find ${\mathcal E}_M \simeq {\rm w} v_1^2
k_\xi(t)/k_0$ since $b_1 \simeq v_1$. Since the kinetic energy
density is dissipated more efficiently than the magnetic one, we
can neglect its contribution in the following discussion. The
time when turbulence is present on scale $\xi_M(t)$ is determined
by Eq.~(\ref{eq:4.7}),  which can be rewritten as
\begin{eqnarray}
\label{taucasBM} \tau_{\rm cas} \simeq \tau_1 \! \left[
\frac{k_0}{k_\xi(t)} \right]^2 \! .
\end{eqnarray}
So instead of considering decaying turbulence, we  consider
stationary turbulence with a scale-dependent duration time (time
during which the magnetic energy is present on that scale),
$\tau_{s1} \simeq \tau_1[k_0/k]^2$ (for $k =k_\xi$ this coincides
with $\tau_{\rm cas}$).

The expression for  ${\mathcal E}_M$  yields the time-independent
magnetic field energy and magnetic helicity power spectra
\begin{equation}
E_M(k,t) = \frac{C_1 v_1^2}{k_0}= \frac12 \, k H_M(k,t), ~~~ k_S <
k < k_0. \label{eq:4.9}
\end{equation}
Here $C_1$ is a constant of order unity, $k_S$ is the smallest
wavenumber where the inverse cascade stops, and the second
equation follows from saturating the causality condition. For the
second stage autocorrelation function, which is inversely
proportional to the turn-over time~(\ref{eq:4.7}), we assume
\begin{eqnarray}
\label{autocorrelationBM} \eta(k) = \frac{\sqrt{2\pi}}{\tau_1}
\left( \frac{k}{k_0} \right)^{\!2} \!,
\end{eqnarray}
to be as close as possible to the first stage description where
$\eta(k_0) = \sqrt{2\pi} /\tau_0$.

At the largest scales there is no efficient dissipation mechanism,
so the inverse cascade will be stopped at the scale $l_S(t) =
2\pi/k_S$ where either the cascade timescale $\tau_{\rm cas}$
reaches the expansion timescale $H^{-1}_\star = H^{-1}(t_\star)$,
or when the characteristic length scale $\xi_M(t) \simeq l_S$
reaches the Hubble radius. These conditions are
$\zeta_\star^{-1/2}l_S^2/v_0 l_0 \leq H^{-1}_\star$ or $l_S \leq
H^{-1} _\star$ (the cascade time is scale dependent and maximal at
$k = k_S$). Defining $\gamma = l_0/ H^{-1}_\star \simeq v_b
H_\star \beta^{-1}$,~\footnote{Taking into account that the
turbulent eddies are formed through bubble expansion and bubble
collisions, the energy-containing scale $l_0$ is determined by
the bubble wall velocity $v_b$ and the expansion time
$\beta^{-1}$ as $l_0 \simeq v_b \beta^{-1}$. We note that
$\gamma$ determines the number ($N_{\rm eddy} \simeq
\gamma^{-3}$) of turbulent eddies within the Hubble radius
$H^{-1}_\star$.}
it is easy to see (taking into account that $v_0$, $\gamma$, and
$\zeta_*$ are less than unity) that the first condition is
fulfilled first and consequently
\begin{eqnarray}
\label{stopBM} \frac{k_0}{k_S} \leq \left( \frac{v_0}{\gamma}
\right)^{\!1/2} \! \zeta_\star^{1/4}.
\end{eqnarray}
To have an inverse cascade requires $k_0/k_S \geq 1$, leading to a
constraint on initial helicity: $\gamma \leq M \zeta_\star^{1/2}$.

\subsubsection{Model B}

Instead of assuming the MHD turbulence model of
Refs.~\cite{BM99,CHB05}, in this subsection we consider  the
alternative MHD turbulence model of
Refs.~\cite{jedamzik,campanelli} (also see Ref.~\cite{son}) with
different evolution laws for the magnetic correlation length and
magnetic and kinetic energy densities,
\begin{eqnarray}
\label{evolution1} && \xi_M(t) \simeq l_0  \! \left(1 +
\frac{t}{\tau_1} \right)^{\!\!2/3} \! , \\
&& \mathcal{E}_M(t) \simeq {\rm w} b_1^2 \! \left(1 +
\frac{t}{\tau_1} \right)^{\!\!-2/3} \! ,
\\
&& \mathcal{E}_v(t) \simeq {\rm w} v_1^2 \!\left(1 +
\frac{t}{\tau_1} \right)^{\!\!-2/3} \! ,
\end{eqnarray}
respectively. Taking into account the above equations, it is easy
to verify, however, that the characteristic turnover and cascade
timescales evolve as in Model A.

It should be noted that, although the magnetic and kinetic
energies scale in time in the same way, equipartition is not
generally reached~\cite{jedamzik}. However, if the initial ratio
of kinetic and magnetic energies is taken to be of order unity,
equipartition is established and lasts for all time. Applying
these results to the case at hand, after the first stage of direct
cascade, at the end of which equipartition between magnetic and
kinetic energy is established, these follows a stage of inverse
cascade which preserves equipartition of energy. Hence we
conclude that kinetic energy cannot be neglected in the second
stage for this particular model of turbulence. In fact the amount
of gravitational radiation produced in the second stage (which, as
we will see, is much greater than that generated in the first
stage) is exactly twice that produced considering only magnetic
energy.

Proceeding as for the discussion for Model A between
Eqs.~(\ref{eq:4.7}) and (\ref{eq:4.9}), we find that the relation
$\mathcal{E}_M \simeq \mathrm{w} v_1^2 \, [k_\xi(t)/k_0]$ does
not change, whereas for the cascade timescale we get
\begin{eqnarray}
\label{taucasBJ} \tau_{\rm cas} \simeq \tau_1 \! \left[
\frac{k_0}{k_\xi(t)} \right]^{3/2} \! .
\end{eqnarray}
The expression for the time-independent magnetic field energy
spectrum does not change, however, now it is valid for $k_S < k <
k_0$ with $k_S$ depending on the particular model of turbulence
adopted (see below).

In Model B, the autocorrelation function is
\begin{eqnarray}
\label{autocorrelationBJ} \eta(k) = \frac{\sqrt{2\pi}}{\tau_1}
\left( \frac{k}{k_0} \right)^{\!3/2} \!,
\end{eqnarray}
while the conditions that determine when the inverse cascade stops
are
$\zeta_\star^{-1/2} l_S^{3/2}/v_0l_0^{1/2} \leq H_\star^{-1}$ or
$l_S \leq H_\star^{-1}$.
However in this case it is easy to see that the former condition
is fulfilled before the latter one, and, consequently, we get
\begin{eqnarray}
\label{stopBJ} \frac{k_0}{k_S} \leq \left( \frac{v_0}{\gamma}
\right)^{\!2/3} \! \zeta_\star^{1/3}.
\end{eqnarray}
The condition to have an inverse cascade, $k_0/k_S \geq 1$, leads
to the same constraint on initial helicity found in Model A, i.e.
$\gamma \leq M \zeta_\star^{1/2}$.

\subsection{Stress-energy tensor and source for gravitational
waves}

The magnetic field perturbation stress-energy tensor is
\begin{equation}
T_{ij}^M ({\mathbf x},t) = {\rm w} b_i({\mathbf x},t) b_j({\mathbf
x}, t).
\end{equation}
For the first decay stage we compute for this magnetic part and
then double the result to account for approximate magnetic and
kinetic energy equipartition for Alfv\'en waves.

To compute the function $H_{ijij}({\bf k},\omega)$, we assume
Millionshchikov quasi-normality~\cite{my75} and adopt the (${\bf
k} \rightarrow 0$) aero-acoustic approximation, which is accurate
for low Mach numbers and slightly overestimates the gravitational
wave amplitude for Mach numbers approaching unity (for details see
Sec.~III of Ref.~\cite{gkk07}). The final result is~\cite{gkk07}
\begin{eqnarray}
H_{ijij}({\bf k}, \omega) \simeq H_{ijij}(0,\omega) = \frac{7
C_K^2 \varepsilon}{6 \pi^{3/2} } \int_{k_0}^{k_d} \!
\frac{dk}{k^6} \exp\!\left( -\frac{\omega^2}{\varepsilon^{2/3}
k^{4/3}} \right)\!{\rm erfc} \!\left(
-\frac{\omega}{\varepsilon^{1/3} k^{2/3}} \right) \! .
\label{eq:4.08bis}
\end{eqnarray}
Here, ${\rm erfc}(x)$ is the complementary error function defined
as $\mbox{erfc}(x) = 1 - \mbox{erf}(x)$, where $\mbox{erf}(x) =
\int_0^x dy \exp(-y^2)$ is the error function~\cite{Gradshteyn}.
The integral in Eq.~(\ref{eq:4.08bis}) is dominated by the large
scale ($k \simeq k_0$) contribution so, for  direct-cascade
turbulence during the first stage, the peak frequency
is~\cite{gkk07}
\begin{equation}
\label{OmegaMaxI} \omega_{\rm max}^{(I)} \simeq k_0 M.
\end{equation}

To compute the gravitational wave source during the second stage
we consider  Model A and Model B separately.

\subsubsection{Model A}

During the second stage, according to Eq.~(\ref{eq:4.06}), kinetic
energy can be neglected compared to magnetic energy. Proceeding
as in the case of Kolmogorov turbulence and making use of the
aero-acoustic approximation, we find
\begin{eqnarray}
H_{ijij}({\bf k}, \omega) \simeq H_{ijij}(0,\omega) = \frac{7
C_1^2 M^3 \zeta_\star^{3/2}\!}{12 \pi^{3/2} k_0} \int_{k_S}^{k_0}
\! \frac{dk}{k^4} \exp\!\left( -\frac{\omega^2 k_0^2}{\zeta_\star
M^2 k^4} \right)\!{\rm erfc} \!\left( - \frac{\omega
k_0}{\zeta_\star^{1/2} M k^2} \right) \!, \label{eq:4.08}
\end{eqnarray}
where $k_S$ can be found by saturating Eq.~(\ref{stopBM}). In this
case the integral is dominated by the  large scale ($k \simeq
k_S$) contribution and is maximal at
\begin{equation}
\label{OmegaMaxBM} \omega_{\rm max}^{(II)} \simeq
\frac{\zeta_\star^{1/2} M k_S^2}{k_0} = 2\pi H_\star \, .
\end{equation}

\subsubsection{Model B}

In Model B of freely decaying MHD turbulence, as discussed in
Sec.\  II.B.2,  during the second stage kinetic energy is
approximatively in equipartition with magnetic energy. To compute
$H_{ijij}({\bf k},\omega)$, we proceed as for the case just
discussed for Model A. The only differences reside in the
expressions for the autocorrelation function and the smallest
wavenumber where the inverse cascade stops, $k_S$ [whose value
can be found by saturating Eq.~(\ref{stopBJ})]. For the case at
hand we find
\begin{eqnarray}
H_{ijij}({\bf k}, \omega) \simeq H_{ijij}(0,\omega) = \frac{7
C_1^2 M^3 \zeta_\star^{3/2}\!}{6\pi^{3/2} k_0^{3/2}}
\int_{k_S}^{k_0} \! \frac{dk}{k^{7/2}} \exp\!\left(
-\frac{\omega^2 k_0}{\zeta_\star M^2 k^3} \right)\!{\rm erfc}
\!\left( - \frac{\omega k_0^{1/2}}{\zeta_\star^{1/2} M k^{3/2}}
\right) \!. \label{H-modelB}
\end{eqnarray}
Again, the integral is dominated by the large scale ($k \simeq
k_S$) contribution and is maximal at
\begin{equation}
\label{OmegaMaxBJ} \omega_{\rm max}^{(II)} \simeq
\frac{\zeta_\star^{1/2} M k_S^{3/2}}{k_0^{1/2}} = 2\pi H_\star \,
.
\end{equation}
The fact that the frequency at the peak of the generated
gravitational wave spectrum is independent of the particular model
of turbulence [compare Eqs.\ (\ref{OmegaMaxBM}) and
(\ref{OmegaMaxBJ})] is easily understood if one observes that it
is the inverse of the maximum cascade timescale $\tau_{\rm
cas}^{\rm max}$, which is $\tau_{\rm cas}^{\rm max} = H_*^{-1}$ by
definition.

\section{Gravitational Wave Spectra}

The total gravitational wave energy density spectrum $\rho_{\rm
GW}(\omega)$ at a given space-time event is obtained by
integrating over all source regions with a light-like separation
from that event, and includes contributions from gravitational
wave generated during the first and second stages. For the first
stage (with duration time $\tau^{(I)}_T = s_0\tau_0$)
$\rho_{GW}^{(I)}(\omega)$ is given by Eqs.~(21) and (A3) of
Ref.~\cite{gkk07}. For the second stage contribution we must
account for the scale dependence of the cascade time. The total
GW fractional energy density parameter at the moment of emission
is~\cite{gkk07}
\begin{eqnarray}
\label{OmegaGW} \Omega_{{\rm GW}}(\omega_\star) \simeq 105 \,
\frac{H_\star^4 \, \omega^3}{H_0^2} \, \sum_m \tau_T^{(m)}
H_{ijij}^{(m)}(0, \omega_\star).
\end{eqnarray}
Here the index $m$ runs over $I$ and $II$ for the first and second
decay stages, $\omega_\star=\omega(t_\star)$ is the angular
frequency of the gravitational wave at the moment of its
emission, and $H_0$ is the current value of the Hubble parameter.

The current gravitational wave amplitude is related to the current
fractional energy density parameter through
\begin{equation}
h_C(f) = 1.26 \times 10^{-18} \left( \frac{\rm Hz}{f} \right)
\left[ h_0^2 \, \Omega_{\rm GW}(f) \right]^{1/2},
\label{gw-amplitude}
\end{equation}
where $h_0$ is the current Hubble parameter in units of 100 ${\rm
km}\,{\rm sec}^{-1}{\rm Mpc}^{-1}$~\cite{m00}. Using the
expressions found for the $H_{ijij}$ tensor, we obtain
\begin{eqnarray}
h_C(f) \simeq 2 \times 10^{-14} \left(\frac{100\,{\rm GeV}}{T_*}
\right) \left(\frac{100}{g_*}\right)^{1/3} \sum_m
\left[\tau_T^{(m)} \omega_\star H_\star^4 H_{ijij}^{(m)}(0,
\omega_\star)\right]^{1/2}. \label{hctoday}
\end{eqnarray}
Here, $f = (a_\star/a_0) f_\star$ is the linear frequency, with
$f_\star = \omega_\star/2\pi$.

%*****************************************   Figure 1  *******************************************%

\begin{figure}[t]
\includegraphics[width=7cm]{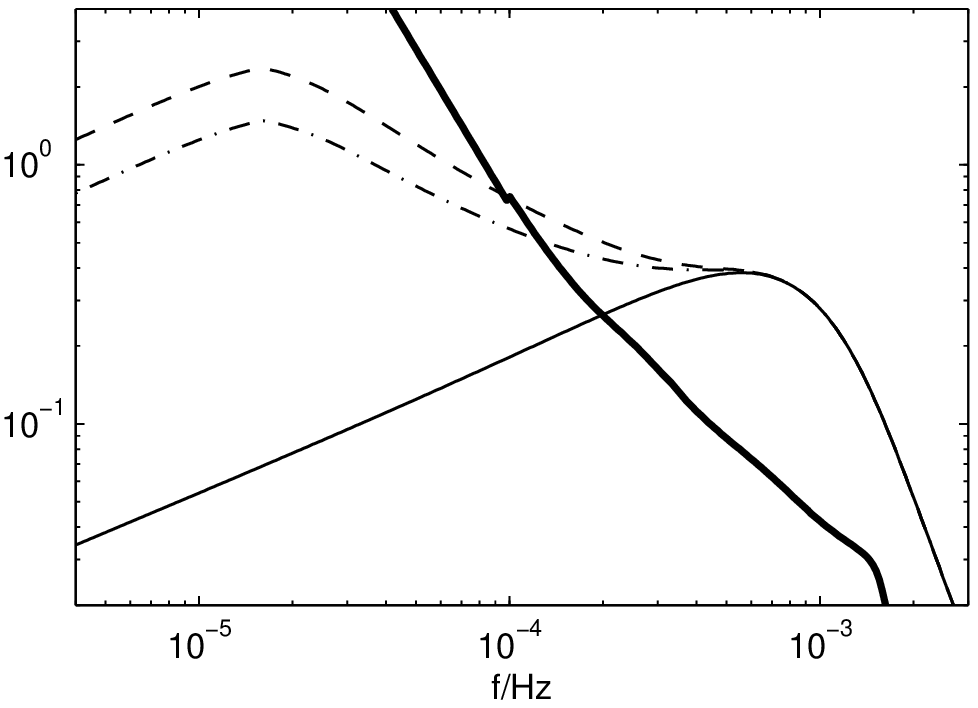}
\hspace{1.7cm}
\includegraphics[width=7cm]{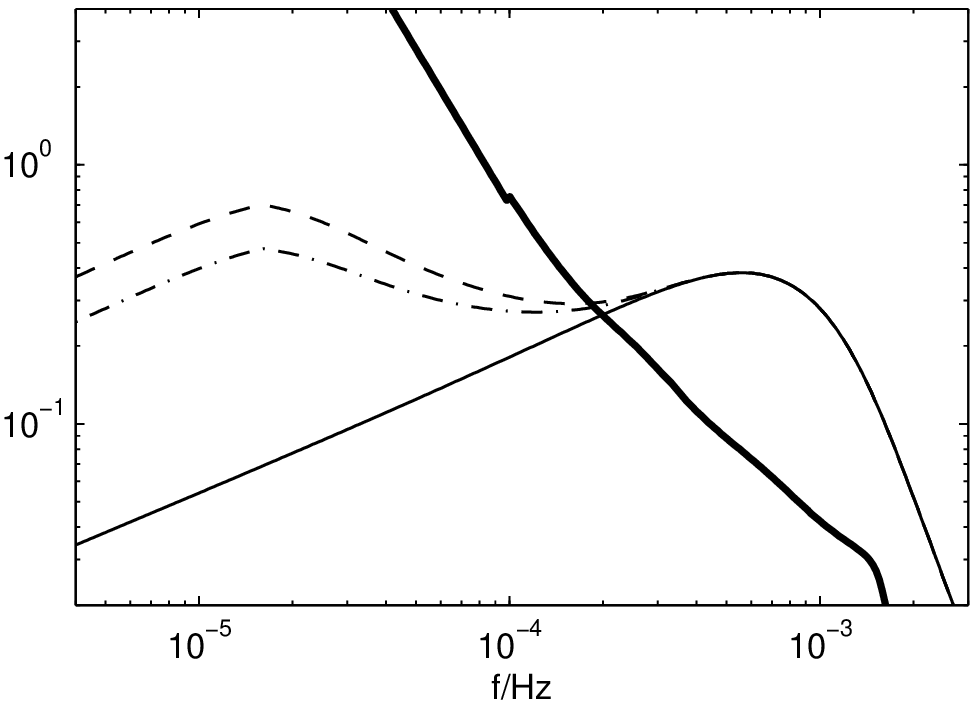}
\caption{The spectrum of the gravitational wave strain amplitude,
$h_C(f)$, as a function of the frequency $f$ for a first-order
phase transition with $g_* = 100$, $T_* = 100$ GeV, $\alpha =
0.5$, and $\beta = 100 H_\star$, from hydrodynamic Kolmogorov
turbulence with zero magnetic helicity (solid lines) and for the
two  MHD turbulence models, Model A (dash-dotted lines) and Model
B (dashed lines). The left panel corresponds to initial magnetic
helicity $\zeta_\star =0.15$, while $\zeta_\star=0.05$ in the
right panel. In both panels the bold solid line corresponds to
the 1-year, $5\sigma$ LISA design sensitivity curve~\cite{curve}
including  confusion noise from white dwarf
binaries~\cite{whitedwarfs}.}
\vspace{0.5cm}
\end{figure}

%*************************************************************************************************%

Figure 1 shows the gravitational wave amplitudes, $h_C(f)$, from
pure hydrodynamical turbulence (no inverse cascade) and for Models
A and B for two different values of initial magnetic helicity,
$\zeta_\star = 0.15$ (left panel) and $\zeta_\star =0.05$ (right
panel). Accounting for inverse-cascade MHD turbulence, the
gravitational wave spectrum has two peaks: the first
higher-frequency one is associated with direct-cascade
hydrodynamical turbulence while the second lower-frequency one is
induced by inverse-cascade MHD turbulence. The amplitude of
gravitational waves emitted during direct-cascade unmagnetized
turbulence peaks at current
frequency~\cite{gkk07,kkgm08,kmk02,dolgov,nicolis}
\begin{equation}
f_{\rm max}^{(I)} = \frac{M k_0}{2\pi} \frac{a_\star}{a_0} =
\left(\frac{M}{v_b}\right) \left(\frac{\beta}{H_\star}\right) f_H.
\label{peakI}
\end{equation}
Here we have used $2\pi/k_0 = l_0 = v_b \beta^{-1}$, where $v_b$
is the bubble wall velocity, and defined
\begin{equation}
f_H = H_\star \frac{a_\star}{a_0} \simeq 1.6 \times 10^{-5}\,{\rm
Hz}\, \left(\frac{g_*}{100}\right)^{1/6}
\left(\frac{T_*}{100\,{\rm GeV}}\right). \label{fh}
\end{equation}
The second peak is at a lower frequency  compared to the
unmagnetized case and is independent of the adopted turbulence
model (since it is determined by the Hubble frequency at the
moment of gravitational wave generation). In fact, the second
peak frequency is
\begin{equation}
f_{\rm max}^{(II)} = f_H . \label{peak}
\end{equation}
Finally, as it is straightforward  to establish, the amplitude of
MHD-turbulence-generated gravitational waves at the peak in Model
A is about a factor $(M/\gamma)^{3/4} \zeta_\star^{9/8}$ larger
than that in the unmagnetized case (and thus strongly depends on
initial magnetic helicity), while it is about a factor $\sqrt{2}
\, (M/\gamma)^{1/12} \zeta_\star^{1/24}$ smaller than that in
Model B.

%*************************************************************************************************%

\begin{figure}[t]
\includegraphics[width=7cm]{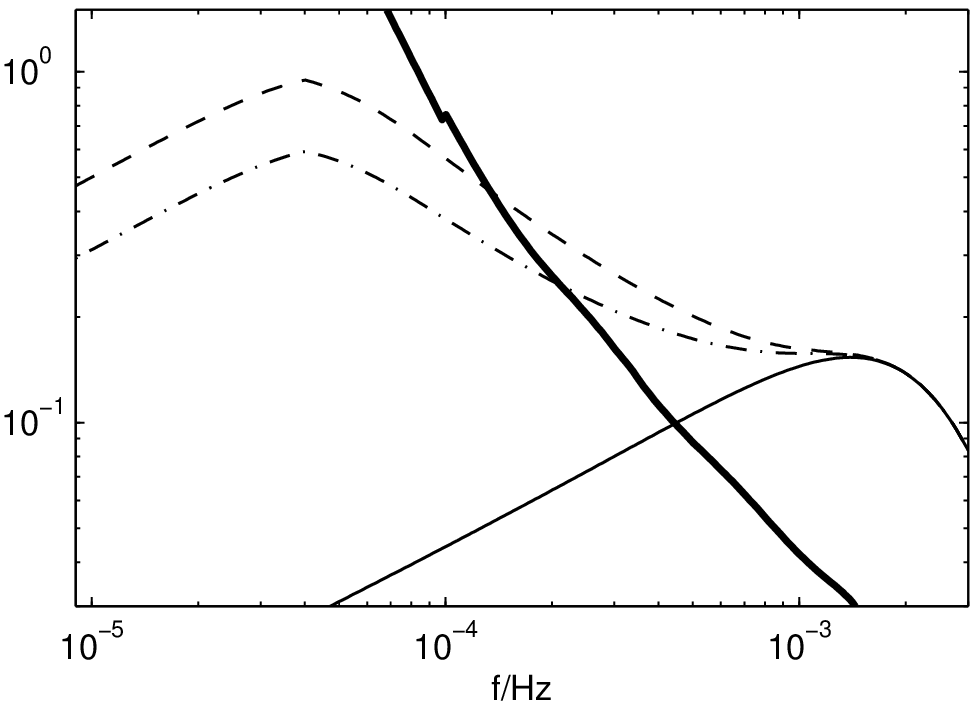}
\hspace{1.7cm}
\includegraphics[width=7cm]{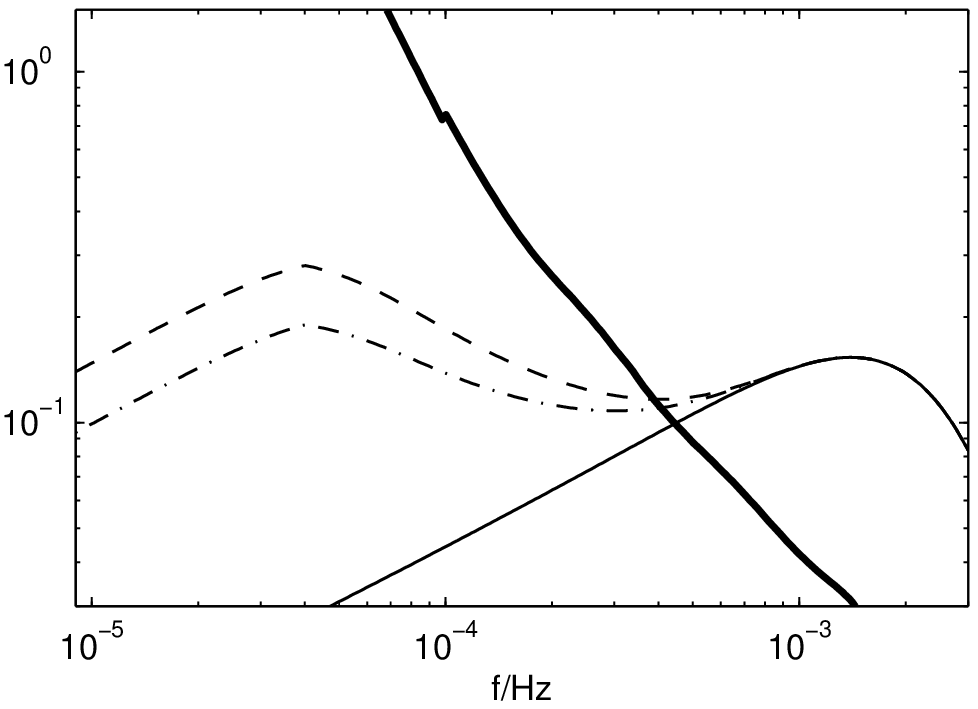} \caption{As in Fig.~1, except now $T_\star =
250$ GeV.}
\end{figure}

%*************************************************************************************************%

In Fig.~2 we show the strain amplitude $h_C(f)$ for a higher phase
transition temperature, $T_\star= 250$ GeV,  with the rest of
the  parameters taking the same values as in Fig.~1. Comparing
Figs.\ 1 and 2 we see that if the phase transition occurs at
higher energy the peak frequency is higher, closer to the LISA
design sensitivity peak.

The gravitational wave signal detection prospects depends
crucially on several parameters: the temperature (the energy
scale) at which the phase transition takes place $T_\star$,
the number of relativistic degree of freedom $g_\star$,
the phase transition model parameters $\alpha$ and $\beta/H_\star$,
and the initial magnetic helicity.

The dependence on $g_\star$ is quite weak, see
Eq.~(\ref{hctoday}). It affects both peak frequencies
(associated with hydrodynamical and MHD turbulence). The initial
magnetic helicity plays a more substantial role if the phase
transition takes places at a higher temperature \cite{kkgm08}. The
phase transition model determines the value of the parameter $\alpha$.
The tensor $H_{ijij}$ appearing in Eq.\ (\ref{hctoday}) depends
on $\alpha$ as well as on the duration time of the direct cascade
stage $\tau_T^{(I)}$ \cite{gkk07}. In addition, the frequency of
the first peak ($\sim k_0 M$) depends on $\alpha$ through the
bubble wall velocity ($v_b$) and other parameters, see Eqs.\
(\ref{kappa})---(\ref{v0}). Increasing $\alpha$ makes the
gravitational wave signal stronger, see Fig.\ 2 of Ref.\ \cite{kkgm08}.
In the context of the detection of gravitational waves, the phase
transition energy scale has two different effects: the peak
frequency is proportional to $T_\star$ but the peak amplitude is
inversely proportional to $T_\star$. These effects must be
accounted for together, see Fig.\ 3 of Ref. \cite{kkgm08}. Another
important parameter is $\beta/H_\star$ which is model dependent
and affects the first peak frequency, see Eq.\ (\ref{peakI}), as
well as the amplitude of the gravitational wave signal. The
dependence on $\beta/H_\star$ can be seen in Fig.\ 5
of Ref.\ \cite{kkgm08}: a larger $\beta/H_\star$
results in a lower amplitude signal with a higher first-peak
frequency, while the position of the second peak,
associated with inverse-cascade turbulence, is not affected.
Figure 3 shows the LISA sensitivity region in the $\beta/H_*$ and
$T_*$ parameter plane for a phase transition with vacuum energy
$\alpha=0.1$ (left panel) and $\alpha=0.5$ (right panel) for Model A.
Figure 4 shows the LISA sensitivity regions in the $\alpha$ and $T_*$ parameter
plane (left panel) and in the $g_\star$ and $T_\star$ parameter plane
(right panel) for Model A.

%%%%%%%%%%%%%%%%%%%%%%%%%%%%%%%%%%%%%%%%%%%%%%%%%%%%%%
\begin{figure}
\includegraphics[width=7cm]{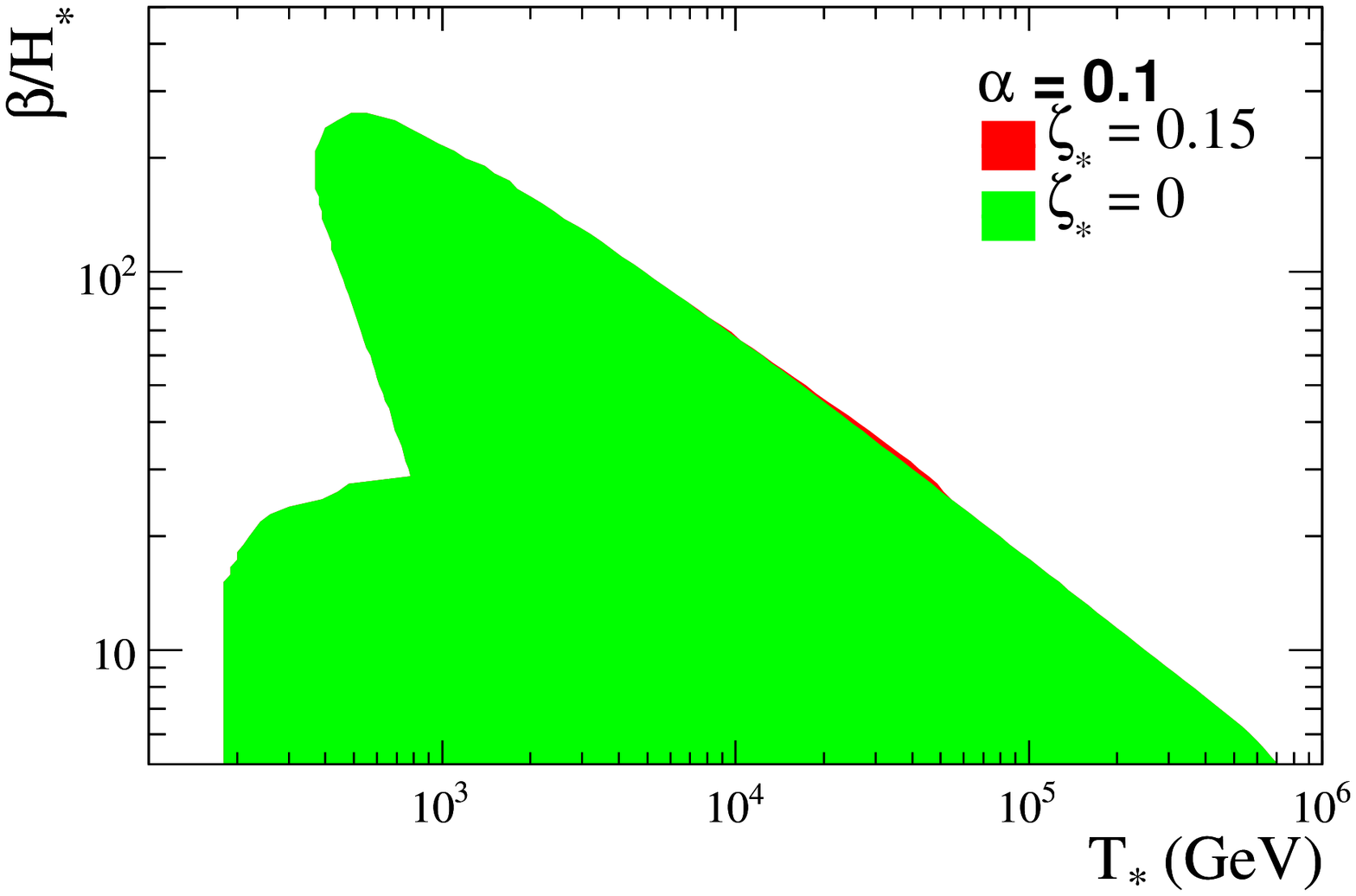}
\hspace{1.7cm}
\includegraphics[width=7cm]{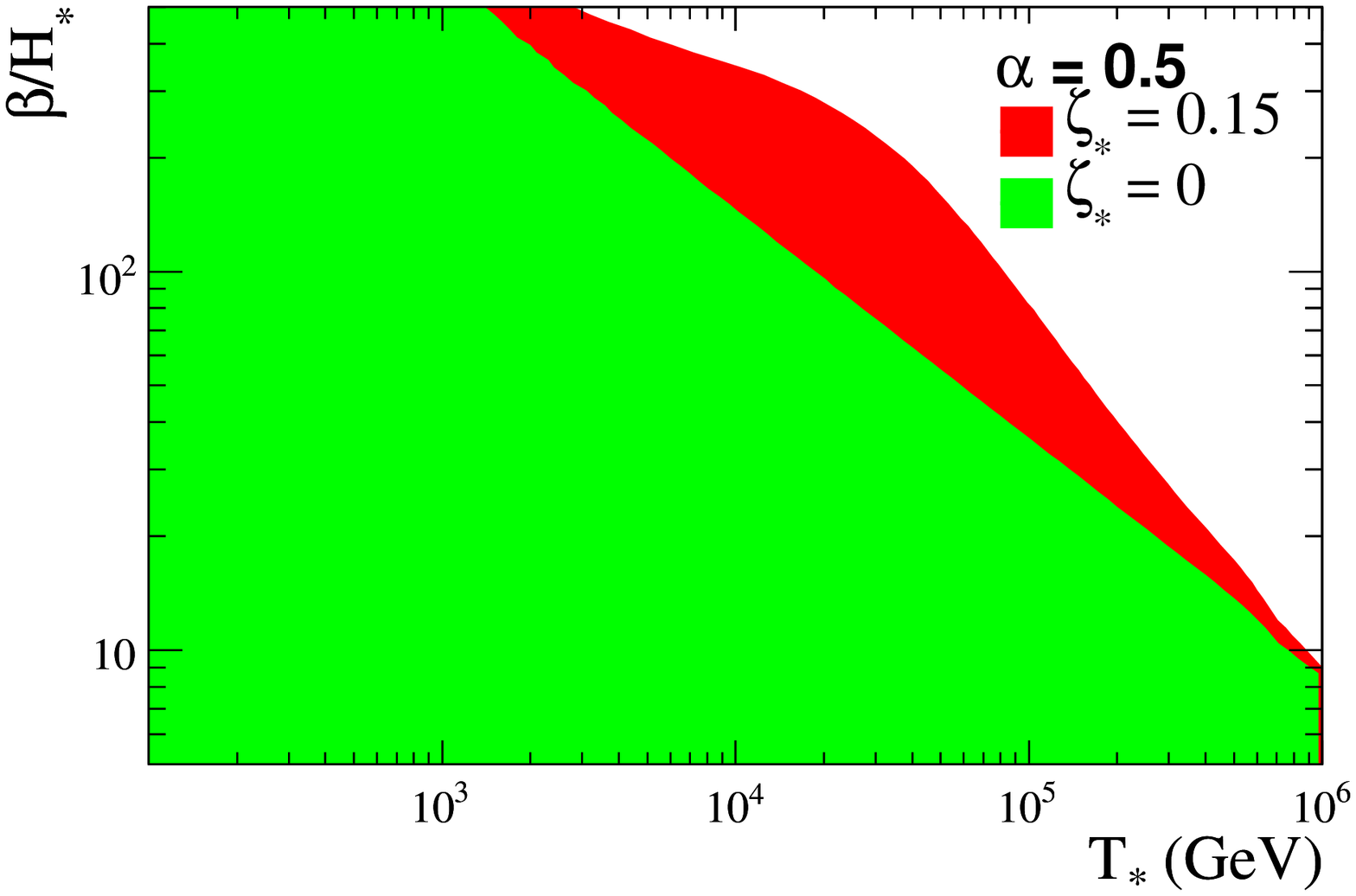}
 \caption{The LISA sensitivity region for Model A in the $\beta/H_*$
and $T_*$ parameter plane for a phase transition with vacuum
energy $\alpha=0.1$ (left panel) and $\alpha=0.5$ (right panel)
\cite{kkgm08}. The regions for $\zeta_*=0$ and $\zeta_*=0.15$
coincide at these temperatures for $\alpha=0.1$ (left panel). A
point in parameter space is considered detectable if at any
frequency its value of $h_c(f)$ is detectable at a
signal-to-noise ratio of 5 in a one-year integration, including
the confusion noise from white dwarf binaries, based on Refs.\
\cite{curve}.}
\end{figure}

%%%%%%%%%%%%%%%%%%%%%%%%%%%%%%%%%%%%%%%%%%%%%%%%%%%%%%
%%%%%%%%%%%%%%%%%%%%%%%%%%%%%%%%%%%%%%%%%%%%%%%%%%%%%%
\begin{figure}
\includegraphics[width=7cm]{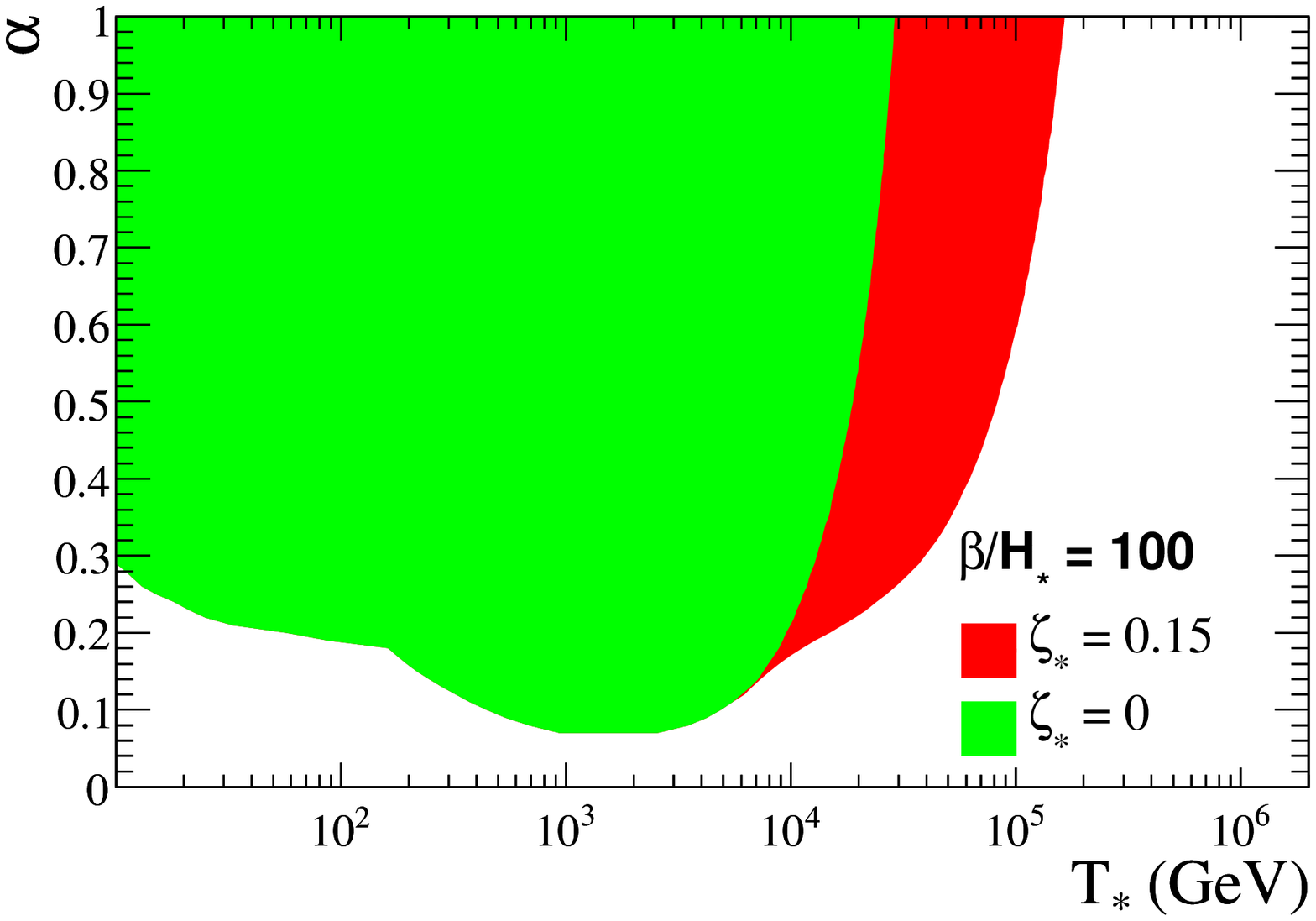}
\hspace{1.7cm}
\includegraphics[width=7cm]{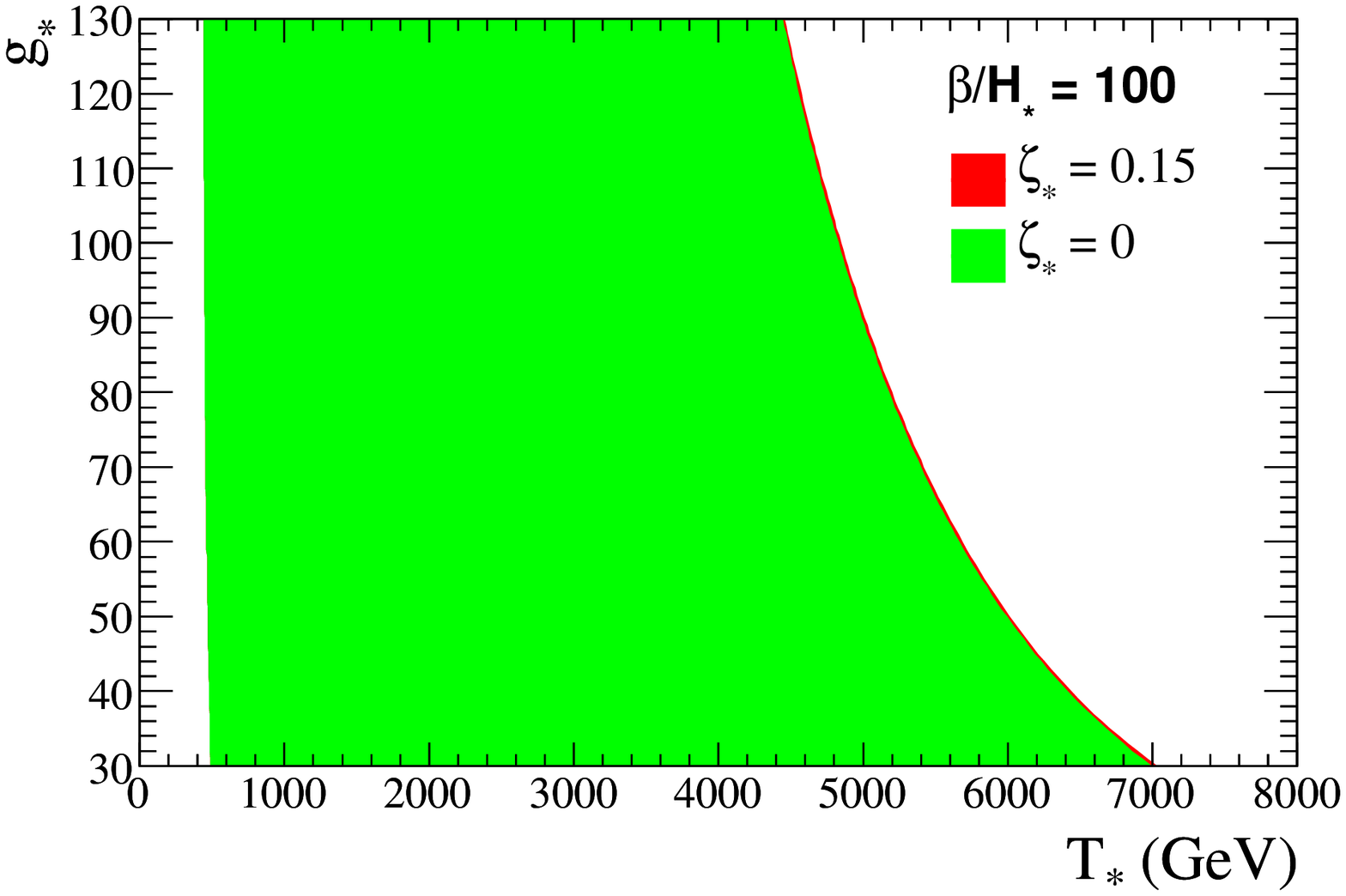}
 \caption{The LISA sensitivity region for Model A in
the $\alpha$ and $T_*$ parameter plane (left panel) with
$\beta/H_\star =100$  and $g_\star =100$, and in the $g_\star$
and $T_\star$ parameter plane (right panel) with $\beta/H_\star =
100$ and $\alpha=0.1$. A point in parameter space is considered
detectable if at any frequency its value of $h_c(f)$ is
detectable at a signal-to-noise ratio of 5 in a one-year
integration, including the confusion noise from white dwarf
binaries, based on Refs.\ \cite{curve}.} \label{fig4}
\end{figure}

%%%%%%%%%%%%%%%%%%%%%%%%%%%%%%%%%%%%%%%%%%%%%%%%%%%%%%
\section{Conclusion}

We have analyzed the generation of gravitational waves at the
electroweak phase transition, from hydrodynamic turbulence in the
presence of a helical magnetic field. It is convenient to
consider this as a two-stage process. During the first stage the
effects of helicity are negligible when studying the production
of gravitational waves from turbulence. The first stage ends when
quasi-conservation of magnetic helicity starts to trigger an
inverse cascade of the magnetic field, that is, a transfer of
magnetic energy from small to large scales. The details of the
inverse cascade process in helical MHD are still under debate and
there exist in the literature two different models. In this paper
we have examined gravitational wave generation in both models of
helical MHD turbulence. We have found that, for realistic values
of the parameters defining the electroweak phase transition, the
generated gravitational wave spectrum  is independent of
turbulence modeling (within a factor of order 2).

The inverse-cascade associated peak frequency always coincides
with the Hubble frequency at the moment of gravitational wave
production. Moreover, as clear from Fig.~1, the main contribution
to the total gravitational waves energy density is from the
second, inverse-cascade stage, even for small values of magnetic
helicity.

Gravitational waves produced {\it via} helical turbulence are
strongly polarized, since magnetic helicity is a parity-odd
quantity and is maximal at the end of the first
stage~\cite{kgr05}. LISA should be able to detect such a
gravitational wave polarization~\cite{seto}. Also, in contrast to
the unmagnetized case, the contribution to the gravitational wave
amplitude coming from the inverse-cascade stage is large enough at
$0.1$ mHz to be detectable by LISA. If the electroweak phase
transition occurs at higher energy (see Fig.~2) the peak frequency
is higher, closer to the LISA sensitivity peak, which leads to a
stronger signal. Our formalism is also applicable to gravitational
wave production at an earlier QCD phase transition, assuming the
presence of colored magnetic fields~\cite{QCD}, or to any other
phase transition~\cite{bubble}: the peak frequency will shift
according to  changes in $T_\star$ and $g_\star$.

Finally, we stress that the gravitational wave signal arising from
 helical MHD turbulent motion exceeds that from bubble
collisions~\cite{kos1,detection,gs06} and that  from
hydrodynamical turbulence alone~\cite{nicolis,gkk07}. Of course,
the strong signal estimated here (see also Ref.~\cite{kgr08})
assumes initial non-zero (although small) magnetic helicity, so
detection of polarized gravitational waves by LISA will indicate
parity violation during the phase transition, as proposed in
Refs.~\cite{helicity}.

\acknowledgments We acknowledge helpful discussions with R.
Durrer, A. Gruzinov, A. Kosowsky, and G. Lavrelashvili. G.G. and
T.K. acknowledge the hospitality of the Abdus Salam International
Center for Theoretical Physics, and support from INTAS grant
061000017-9258 and Georgian NSF grants ST06/4-096 and ST07/4-193.
T.K., Y.M., and B.R. acknowledge US DOE grant DE-FG03-99EP41093.

\end{document}